\documentclass{article}
\usepackage{graphicx}
\usepackage{amsmath}
\usepackage{graphicx}
\usepackage{subcaption}
\usepackage{hyperref}
\usepackage[a4paper,margin=1in]{geometry}

\newcommand{\UAMI}{\operatorname{UAMI}}
\newcommand*{\pout}{p_\mathrm{out}}
\newcommand*{\pin}{p_\mathrm{in}}

\title{A model for generating temporal networks with dynamic community structure guided by mutual information}
\author{Peijie Zhong, Ra\'{u}l Mondrag\'{o}n and Richard Clegg}
\date{Dec, 2025}
\begin{document}

\maketitle

This paper introduces a generative model for temporal networks that jointly controls community evolution and dynamic node sets. The model represents community structure as a sequence of partitions and uses a genetic search guided by a similarity measure based on mutual information to regulate changes between snapshots. This allows explicit control of community evolution including splits and merges while handling node additions and removals. Temporal edges are then generated using intra- and inter-community probabilities derived from data or theoretical bounds to ensure connectivity. Simulation experiments on real-world datasets demonstrate the ability of the generative model to model the evolution of real dynamic communities. The model is used as a benchmark to study the impact of the rate at which nodes join/leave the network on the performance of dynamic community detection algorithms.

\section{Introduction}
Communities are a defining mesoscopic feature of many temporal systems, ranging from communication and financial infrastructures (\cite{li2022multilayer, kamuhanda2023illegal, hassan2022anomaly}) to online platforms~\cite{lin2008facetnet, mohammadmosaferi2020evolution,  tajeuna2015tracking}, whose interaction patterns evolve continuously. While communities in static networks have been widely studied, dynamic communities in temporal networks have received less attention. Generative models for dynamic communities are not common but are important. It is essential for a generative model for temporal networks to include communities if it is to be close to real world data. A generative model can serve as a null model to help understand the features in real network data~\cite{bazzi2020framework}. A generated network with known ground truth communities is vital to test algorithms that detect dynamic communities in temporal networks~\cite{bazzi2016community, mucha2010community, pamfil2019relating, dabideen2014clan, aslak2018constrained, kuncheva2017restricted, jiao2017temporal}. Temporal networks with a realistic structure can also be important to benchmark graph processing systems~\cite{huang2023temporal,steer2023raphtory}.

While there are existing generative models for temporal networks with dynamic communities~\cite{bazzi2020framework, karrer2011stochastic}, this work has significant advantages. Our model explicitly captures the fact that communities split, merge, grow, and contract, reshaping the organisational structure of the network over time. Existing generators assume a fixed node set where in reality often nodes leave a network or join after the initial time period. Our model allows the explicit tuning of how quickly communities change in time rather than this being indirectly controlled by parameters. In addition, our model can be explicitly tuned to fit real data. Furthermore, this model provides link stream data where links have timestamps, this is the form many graph processing system need to ingest data.  

Our new model simultaneously captures the rate of community evolution across time and captures explicit community events (splits/merges) and node churn (joining/leaving). The first stage is to represent the network's community structure as an ordered sequence of partitions and to search this space with a genetic algorithm guided by a similarity measure we call union adjusted mutual information~\cite{zhong2024quantifying} designed for tackling the situation where the node set changes across time steps. In the second stage, the model inserts timestamped edges within each time step based on the community labels generated by the genetic algorithm. This process follows the principle of the stochastic block model (SBM): the probability of forming a temporal edge between a pair of nodes depends on their community labels. Unlike traditional SBM-based generators that produce static graphs, our model assigns timestamps to edges, that produce a temporal graph at each time step. The edge-generation parameters can be set to ensure the network is connected and has easy to detect communities or, if the goal is to model existing data, inferred from the parameters of the target network. 

We validate the framework on diverse real-world temporal networks, including a network of emails, a network of citations among physics papers, a network of social platform interaction and a network of non-fungible token (NFT) transactions. The results show that it reproduces dynamical characteristics of community evolution. Furthermore, the model provides a benchmark for studying the impact of the rate at which the study nodes joining/leaving the network on the performance of the dynamic community detection algorithm.
\section{Related work}
Current research on network generation is widely applied to tasks such as network simulation, construction of synthetic benchmarks, and time modelling. Among them, the community structure, as a core element for characterising topological features, has always been a key focus of modelling. In a static situation, the Stochastic Block Model (SBM)~\cite{holland1983stochastic} uses a prior community partition and assigns differentiated edge connection probabilities within and between communities, forming the basic paradigm for studying detectability~\cite{abbe2015community, abbe2018community} and inference limits. Although some of its extensions have enhanced the expressiveness in terms of degree heterogeneity~\cite{lancichinetti2008benchmark, lancichinetti2009benchmarks} and mixed membership~\cite{airoldi2008mixed}, they still remain within the framework of static graph and fixed node set in general, and thus are unable to directly be applied to controllable dynamic generation scenarios.

In parallel, extensive progress has been made in dynamic community detection, with numerous algorithms proposed to detect evolving community structures from temporal graphs~\cite{huang2021survey}. However, despite these advances in detection methods, the development of controllable generative models for dynamic networks remains relatively limited. In recent years, dynamic networks, especially generation models for constructing dynamic networks with dynamic community structures, have attracted a great deal of attention. There are mainly three types of paradigm for the generative modelling of dynamic networks. 

The first is probabilistic graph models that explicitly depict the evolution process of communities along the time dimension. The dynamic infinite relational model (dIRM) can express events such as the birth, death, splitting, and merging of communities without predefining the number of groups~\cite{ishiguro2010dynamic}. The advantage of these methods lies in the ability to incorporate complex evolutions into a unified statistical inference framework. However, it focusses on the estimation and interpretation of parameters and posteriors, rather than generating controllable results under the premise of a given temporal community similarity trajectory. Moreover, the model does not directly model the joining and leaving of nodes but instead uses a cluster to include the inactive nodes. Building on the same paradigm, the dynamical degree-corrected stochastic block model (D-DCSBM)~\cite{wilson2019modeling, dall2020community} assumes Markovian persistence of community labels and generates each snapshot via a degree-corrected SBM, which supports analyses of detectability thresholds and scalable spectral algorithms, yet it the model treats the absence nodes in some time steps as  implicitly isolated (degree-zero) nodes, rather than modelling explicit joining/leaving. Although it is feasible to model the nodes that are missing at certain time steps as degree-zero nodes in some downstream tasks~\cite{dall2020community}, for a general generative model, criticism points out that retaining a large number of absent nodes as isolated points will change the spectral properties~\cite{von2007tutorial} of the graph and degrade the representation and community detection based on spectral/random walks: the multiplicity of zero eigenvalues of the graph Laplacian matrix is equal to the number of connected components. Adding isolated nodes directly increases the multiplicity of zero eigenvalues and dilutes the informative low-frequency spectral structure. Moreover, simply treating the departed nodes as degree-zero nodes confuses structural absence and behavioural silence~\cite{kossinets2006effects, krivitsky2011adjusting}, causing statistical bias and making indicators such as density and average degree incomparable.

Another important thread is the mesoscale structure synthesis framework for multi-layer networks, which treats dynamic networks as ordered multi-layer networks. This framework consists of two steps: inter-layer label inheritance/perturbation and intra-layer edge connection based on labels, to construct cross-layer changing community structures, thereby providing a standardised generation idea for the evaluation of dynamic community detection~\cite{bazzi2020framework}. Although this framework introduces the indirect shaping of inter-layer similarity by introducing the probability and null distribution of inherited community labels, it does not guarantee to achieve a specified inter-layer similarity target.

The data-driven dynamic network generation model directly generates timestamp edges at the event level, for example, by learning interaction distributions through time random walks~\cite{almasan2025generating, gupta2022tigger} or subgraph~\cite{zeno2021dymond} decompositions, and then outputs time series using auto-regressive or point process mechanisms~\cite{zhou2020data, gupta2022tigger}. This approach demonstrates strong capabilities in matching empirical statistics, but usually relies on training data to learn distributions and has weak interpretability and controllability constraints for community evolution, making it difficult to serve as a tunable benchmark for studying the relationship among community similarity and nodes' joining/leaving.

Compared with the above work, this paper, under the premise of allowing the addition and deletion of nodes and the splitting/merging of communities, introduces the union-adjusted-mutual-information (UAMI) that can handle variable node sets as a cross-time similarity metric and incorporates it into the genetic search process to explicitly and directly model the community similarity between adjacent snapshots. This design unifies the interpretability of mesoscopic label inheritance, intra-layer edge based on labels and the time fidelity of event-level timestamp generation: on the one hand, we directly generate edges with timestamps within each time step, using the Poisson process as the baseline (which can be replaced by more complex self-exciting point processes) to describe the waiting time and arrival rate; on the other hand, the edge formation probability is determined by the community pairs, thereby achieving controllability in both who is connected with whom and when to connect. 
\section{Problem and Method}
A dynamic network refers to a network in which the interaction relationships between nodes change over time. Common dynamic network representation methods can be classified into two categories. The snapshot model divides the time dimension into a series of discrete time windows, and constructs a series of static graphs $\mathcal{G}=\left(G^{(1)},\dots, G^{(T)}\right)$ to represent the snapshot in each time window, where $G^{(t)}=(V_t,E_t)$, $V_t$ represents the nodes that involved in at least one interaction in the $t$-th time window. The link stream model records each interaction as a timestamped edge $(u,v,t)$, which can accurately reflect the time information of the event. In this work, we aim to generate a temporal network that combines the advantages of both representations. Specifically, we generate a link stream model in which the temporal edges are derived from snapshot-based community structures. 

We represent the dynamic community structure of a temporal network as an ordered sequence of partitions $\left(\mathcal{C}^{(1)}, \dots, \mathcal{C}^{(T)}\right)$, where $T$ is the last time step. At time $t$, the partition is $\mathcal{C}^{(t)}=\{C^{(t)}_1, \dots, C^{(t)}_{k^{(t)}}\}$, covering the node set $V_t$, where $C^{(t)}_i$ is the set of nodes in community $i$ during window $t$ and $k^{(t)}$ represents the number of communities at time step $t$. Note that different partitions may involve different node sets, so $V_i$ and $V_j$ do not need to be equal.

We adopt a two-step method to generate the temporal network. Take generating a dynamic network with community structure with $T$ time steps as an example. In the first step, the genetic algorithm accepts the initial community partition $\mathcal{C}^{(1)}$ and the sequence of community structure similarity between adjacent time steps $\left(s^{(1,2)},\dots,s^{(T-1,T)}\right)$ as input. In this work, we employ a measurement~\cite{zhong2024quantifying} based on mutual information to measure the similarity of community structures. The model searches for the partition that has a given similarity to the community partition input in the previous time step, thereby constructing a temporal partition with $T$ time steps. In the second step, the model inserts edges with timestamps one by one based on $p^{(t)}_{\mathrm{in}}$ and $p^{(t)}_{\mathrm{out}}$. These parameters control the density of edges intra- and inter-communities within the stochastic block model (SBM) at time window $t$, and will be formally defined later in Section~\ref{sec:insert-link}. The same pair of nodes may be connected by multiple temporal edges, reflecting repeated interactions over time. 
\subsection{The initial partition}
At the very first time step ($t=1$), no prior community labels are available, so we begin by assuming some initial number of communities and sizes. Nodes are assigned at random to these communities with probability proportional to the size. This is equivalent to drawing from a Dirichlet distribution. Let $k^{(1)}$ be the number of communities at the first time step. We draw the community-mixing proportions $\pi=(\pi_1,\dots,\pi_{k^{(1)}})\sim \mathrm{Dir}(\alpha)$ with concentration parameters $\alpha=(\alpha_1,\dots,\alpha_{k^{(1)}})$, $\alpha_j>0$. Given $\pi$, each node $i$ is independently assigned to community $j$ with probability $\Pr(c_i^{(1)}=j)=\pi_j$.
\subsection{Genetic algorithm framework}
\label{sec:genetic}
\begin{figure}
    \centering
    \includegraphics[width=0.8\linewidth]{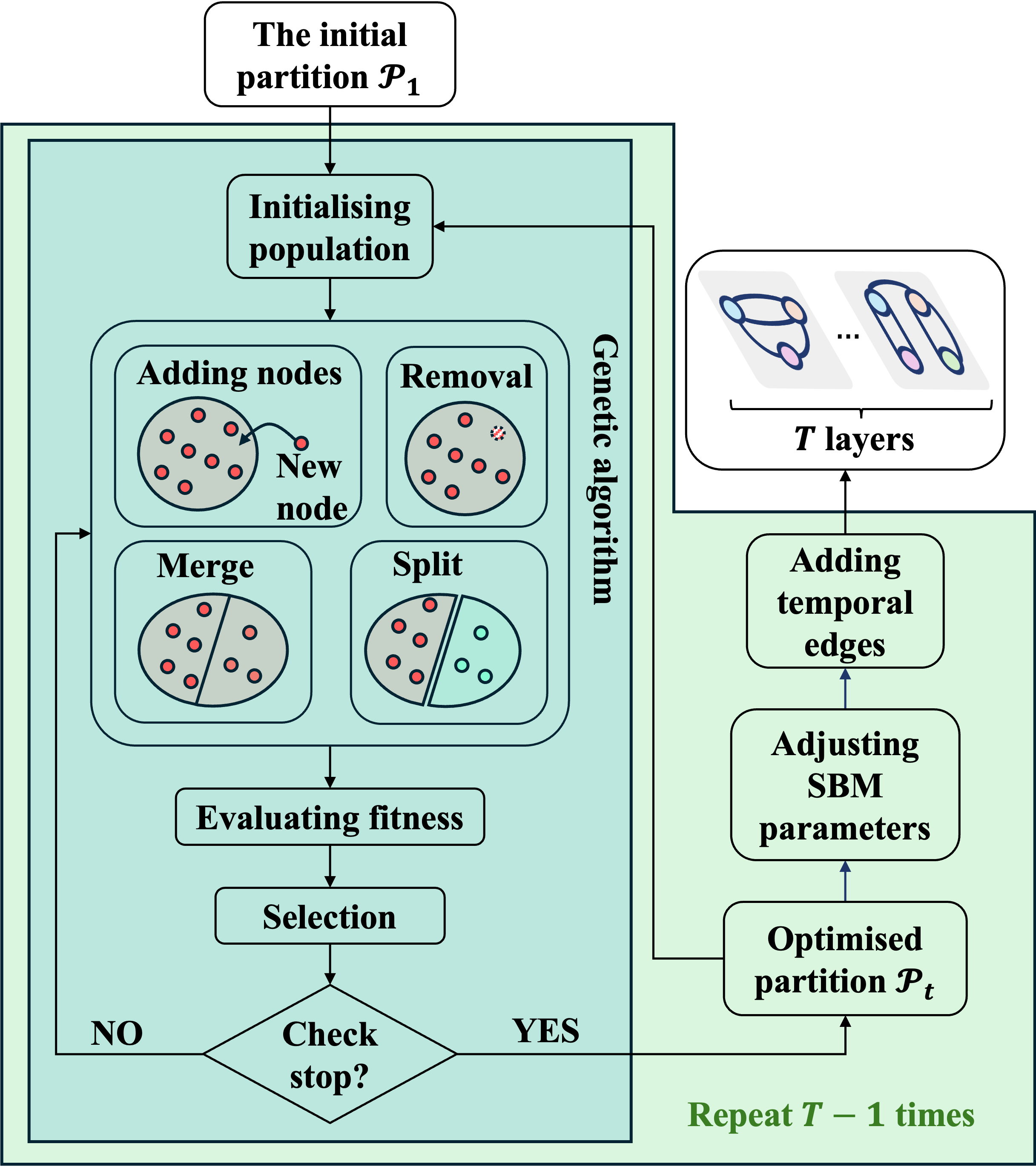}
    \caption{The figure demonstrate the work flow of the generative model: the first step is applying a genetic algorithm to generate dynamic partitions, and the second step is inserting temporal edges based on the generated partitions. }
    \label{fig:flow-chart}
\end{figure}
The first step of the model is to generate community labels for each node at every time step. To control the rate of structural change while accounting for an evolving node set, we employ a genetic algorithm. By simulating key community events, such as node addition and removal, and community splits and merges, the algorithm efficiently explores the solution space to produce temporally consistent dynamic community structure. The blue box in Figure~\ref{fig:flow-chart} illustrates the process of genetic algorithm. 
\subsubsection{Initialisation}
We first introduce the notations used in the initialisation stage. Let $N_{\text{pool}}$ denote the maximum number of candidate nodes, including both existing nodes and nodes that may appear in future time steps. At time step $t$, the current community partition is denoted by $\mathcal{C}^{(t)}$. The genetic algorithm receives the community partition at time step $t-1$ as the basis for searching for the community partition at time step $t$. We denote the solution set obtained by the genetic algorithm after iterating up to the $j$-th round as $S^{(t)}_j=\{\mathcal{C}^{(t)}_{1,j}, \dots, \mathcal{C}^{(t)}_{r,j}\}$, where $r$ is the size of population of the genetic algorithm. The method of constructing the initial solution set $S_1^{(t)}$ is to copy the partition of the $t-1$ time step $r$ times, and then apply a perturbation to each copy. Specifically, for every node that is present, its community label is reassigned with a small probability, where the new label is sampled uniformly from $\{1, \dots, k^{(t-1)}\}$. This strategy preserves the structural information of the current partition, while introducing controlled randomness to increase population diversity, which helps the genetic algorithm explore the solution space more effectively.
\subsubsection{Mutation}
Mutation introduces random modifications to a community partition in order to enhance the exploration ability of the genetic algorithm. In dynamic networks, community structures typically evolve through several fundamental events. We therefore design the mutation operators to align with these basic community evolution events, ensuring that the search process is both flexible and semantically meaningful. Specifically, we consider four types of mutation operations:
\begin{itemize}
    \item Adding nodes: Introduce nodes that are currently unassigned from the node pool. If a node has appeared in previous time steps, it may return to its original community with a certain probability; otherwise, it is assigned to a randomly selected community.
    \item Removing nodes: Remove nodes from the current partition by marking them as unassigned. 
    \item Splitting a community: Select a community and divide its nodes into two disjoint subsets. One subset retains the original label, while the other is assigned a newly created label, forming a new community. 
    \item Merge communities: Select two communities and assign the same label to all their nodes, effectively combining them into a single community.
\end{itemize}

These four operators collectively capture the principal modes of community evolution in dynamic networks. To further control the addition and removal of nodes, we adaptively adjust their probabilities based on the difference in node count between the current candidate partition and the desired partition at time step $t$. Let $\mathcal{C}_{i,j}^{(t)}$ be the $i$-th candidate community partition produced by the genetic algorithm at the $j$-th iteration, and $n^{(t)}_{i,j}$ denotes its number of assigned nodes. Suppose the target number of nodes for step $t$ is $\hat{n}^{(t)}$. Then, for nodes that are assigned community labels, they will be removed with probability $p_{\mathrm{remove}}=\max\left(0,\frac{\hat{n}^{(t)}-n^{(t)}_{i,j}}{\hat{n}^{(t)}}\right)$. Conversely, for nodes that are not assigned community labels, they will be added into the partition and being assigned community labels with probability $p_{\mathrm{add}}=\max\left(0,\frac{n^{(t)}_{i,j}-\hat{n}^{(t)}}{\hat{n}^{(t)}}\right)$. This adaptive mechanism ensures that the mutation operators adjust the node set toward the desired size while maintaining flexibility in exploring the search space.
\subsubsection{Fitness function}
The genetic algorithm produces a set of candidate community partitions at the $j-$th iteration $S^{(t)}_j=\{\mathcal{C}^{(t)}_{1,j}, \dots, \mathcal{C}^{(t)}_{r,j}\}$. We evaluate each candidate using a fitness function that balances two objectives. The first is maintaining a prescribed level of similarity to the input partition at time $t-1$, and the second is producing a partition with the desired number of nodes. Let $s^{(t-1,t)}$ denote the expected similarity between time step $t-1$ and $t$. To measure the similarity between partitions defined on possibly different node sets, we adopt the union-adjusted mutual information (UAMI)~\cite{zhong2024quantifying}, an extension of the mutual information index. UAMI takes values in $[0,1]$, where higher values indicate more similar partitions. The fitness value of the $i$-th candidate community partition at the $j$-th iteration is defined as follows:
\begin{equation*}
   1-\left(\left|\UAMI\left(\mathcal{C}_{i,j}^{(t)},\mathcal{C}^{(t-1)}\right)-s^{(t-1,t)}\right|+\lambda\frac{|n_{i,j}^{(t)}-\hat{n}^{(t)}|}{\hat{n}^{(t)}}\right).
\end{equation*}
The scalar $\lambda \geq 0$ balances these two objectives. By construction, the fitness value lies in $(-1,1]$ when $\lambda=1$. A higher fitness value indicates better agreement with both the desired similarity and the target node membership, thus guiding the genetic algorithm toward partitions that evolve at the prescribed rate while faithfully reflecting node addition and removal.
\subsection{Constructing temporal network from a partition}
\label{sec:insert-link}
Since the community structure of the network is given, we can draw on the idea of the stochastic block model to construct the temporal network. In time step $t$, given the probability that two nodes are connected: 
\begin{equation*}
    P^{(t)}_{u,v}=
    \begin{cases}
        p^{(t)}_{\mathrm{in}} \quad \text{if } c^{(t)}_u=c^{(t)}_v,\\
        p^{(t)}_{\mathrm{out}}\quad \text{otherwise},
    \end{cases}
\end{equation*}
where $c^{(t)}_u$ and $c^{(t)}_v$ represents the community label of node $u$ and node $v$ at the $t$-th time step. When simulating empirical data, the probability can be inferred directly from the densities observed within and between communities~\cite{peixoto2018nonparametric, peixoto2023descriptive} (see section~\ref{sec:simulation}). However, in purely synthetic settings these parameters are not available a priori (section~\ref{sec:test}). Randomly choosing $\pin^{(t)}$ or $\pout^{(t)}$ might disrupt the network's connectivity. For example, a too low value of $\pout^{(t)}$ will makes the network broken between communities at that time step, meanwhile, $\pin^{(t)}$ needs to be sufficiently higher than $\pout^{(t)}$ to ensure that communities are detectable. To address this, we derive the lower bound for $\pout^{(t)}$ that guarantees the global connectivity with high probability, enforced by choosing $\pout^{(t)}$ so that the probability of any pair of communities having zero inter-edges is below a specified tolerance $\epsilon$. Given the condition of $\pout^{(t)}$, the selection of $\pin^{(t)}$ becomes more flexible. For instance, $\pin^{(t)}$ can be calculated based on the detectable limit provided in the paper~\cite{abbe2015community, abbe2018community}.

Through the above genetic algorithm, we can obtain series partitions of node sets, denoted as $(\mathcal{C}_1, \dots, \mathcal{C}_T)$, where $\mathcal{C}_t=\left\{C^{(t)}_1 ,C^{(t)}_2, \dots, C^{(t)}_{k^{(t)}}\right\}$ represents the community structure at time step $t$. Therefore, we need to insert edges with timestamps to transform them into a temporal network. We denote the number of nodes in the $i$-th community as $n_i^{(t)}$, then, for any two communities $C^{(t)}_x$ and $C^{(t)}_y$, the number of possible edges between them is $n^{(t)}_x n^{(t)}_y$. Then the probability that community $C^{(t)}_x$ and community $C^{(t)}_y$ is separated is $(1-p^{(t)}_{\mathrm{out}})^{n^{(t)}_x n^{(t)}_y}$. Let us use $B^{(t)}_{xy}$ to represent the event that there are no edges between community $C^{(t)}_x$ and community $C^{(t)}_y$. Then the probability that there is at least one pair of communities where there is no edges between them is $P \left( \bigcup_{x<y} B^{(t)}_{xy}\right)$. According to Boole's inequality, we have
\begin{equation*}
    P\left(\bigcup_{i<j} B^{(t)}_{ij}\right)\leq \sum_{i<j} P(B^{(t)}_{ij}).
\end{equation*}

We use Taylor expansion to approximate $P(B^{(t)}_{ij})=(1-p^{(t)}_{\mathrm{out}})^{n^{(t)}_i n^{(t)}_j}$. Since the negative exponential function is monotonically decreasing, and the value of the sum $\sum_{z=1}^\infty {\left(p_{\mathrm{out}}^{(t)}\right)}^z/z$ increases as the number of expanded terms increases, we have inequalities:
\begin{equation*}
    \begin{aligned}
        P(B^{(t)}_{ij})& =(1-p_{\mathrm{out}}^{(t)})^{n_i^{(t)}n_j^{(t)}}=e^{n_i^{(t)}n_j^{(t)}\ln(1-
        \pout^{(t)})}\\
        &=\exp\left(-n_i^{(t)}n_j^{(t)}\left(\sum_{z=1}^\infty\frac{{p_{\mathrm{out}}^{(t)}}^z}{z}\right)\right)\\
        &< \exp\left(-n_i^{(t)}n_j^{(t)}p^{(t)}_{\mathrm{out}}\right).
    \end{aligned}
\end{equation*}

We denote $o^{(t)}=\mathop{\min}\limits_{i \neq j} \{n_i^{(t)}n_j^{(t)}\}$. Then, every pair of communities in $\sum_{i<j} P(B^{(t)}_{ij})$ is not larger than $\exp(-o^{(t)}
\pout^{(t)})$, so we have that the upper bound of $\sum_{i<j}P(B^{(t)}_{i j})$ is
\begin{equation*}
    \sum_{x<y} P({B^{(t)}_{xy}}) \leq \frac{k^{(t)}(k^{(t)}-1)}{2}\exp(-o^{(t)}p^{(t)}_{\mathrm{out}}).
\end{equation*}
Let $\epsilon$ be a sufficiently small probability that is the maximum probability that we can tolerate the disconnecting of two communities. We require $\binom{k^{(t)}}{2}\exp(-o^{(t)}\pout^{(t)})<\epsilon$, then we have $\pout^{(t)} > \frac{1}{o^{(t)}}\left(\ln{\binom{k^{(t)}}{2}}-\ln \epsilon \right)$. In extreme cases, there may be small-sized communities in the network. In this case, $\pout^{(t)}$ will be relatively large, even exceeding 1. Therefore, we need to set the maximum value of pout so as to ignore those smaller communities. Given that $\pout^{(t)}<(\pout^{(t)}
)_{\max}$, where $(\pout^{(t)})_{\max} \in (0,1)$, we should ignore the communities whose size is smaller than $\left\lceil \sqrt{\ln{\binom{k^{(t)}}{2}}-\ln \epsilon/\left(\pout^{(t)}\right)_{\max}} \right\rceil$.

To generate temporal edges with realistic temporal patterns, each inserted edge in time step $t$ is assigned a timestamp sampled from a Poisson process. Specifically, for any potential edge $(u,v)$ that is formed with probability $P_{u,v}^{(t)}$, we draw its occurrence times $\{\tau^{(t)}_{e_1},\tau^{(t)}_{e_2},\dots\}$ from a Poisson distribution with rate parameter $\gamma^{(t)}$, where $\tau_{e_i}^{(t)}$ is the timestamp of the $i$-th edge in the $t$-th time step. The rate controls the expected number of interactions in unit time within the $t$-th time step.

The mathematical derivation of the lower bound for $p_{out}^{(t)}$ serves a critical physical purpose in the context of realistic network modelling. In many real-world systems, such as communication networks or citation graphs, the network typically maintains a giant connected component that encompasses the majority of nodes. A fragmented network consisting of many small, isolated islands is often a pathological case or indicates a system in its very early infancy. Besides, from the angle of community detection algorithm, if the inter-community connection probability $p_{out}^{(t)}$ falls below the derived threshold, the generated snapshot risks shattering into disconnected components corresponding to the individual communities. From a spectral graph theory perspective, such disconnection leads to a multiplicity of zero eigenvalues in the graph Laplacian matrix. This is not merely a theoretical issue, it artificially inflates the difficulty for spectral clustering algorithms and random-walk-based methods, which rely on the spectral gap to detect community boundaries.
\section{Results}
In this section, we evaluate the performance and utility of the proposed generative model from four complementary perspectives. First, we validate the quality of the synthesised networks by analysing their temporal community strength using multilayer modularity. Second, we demonstrate the model’s capacity to simulate empirical data by reproducing the community evolution patterns of four diverse real-world temporal networks, discussing its role as a memoryless null model. Third, we deploy the model as a controllable benchmark to systematically assess the robustness of state-of-the-art dynamic community detection algorithms against varying rates of node churn and structural evolution. Finally, we provide empirical evidence of the genetic algorithm’s convergence, ensuring it reliably satisfies the prescribed constraints on community similarity and node population size.
\subsection{Quality of community structure}
\begin{figure}[t]
\centering
\subcaptionbox{Intra-layer modularity\label{fig:modularity}}{\includegraphics[width=\linewidth]{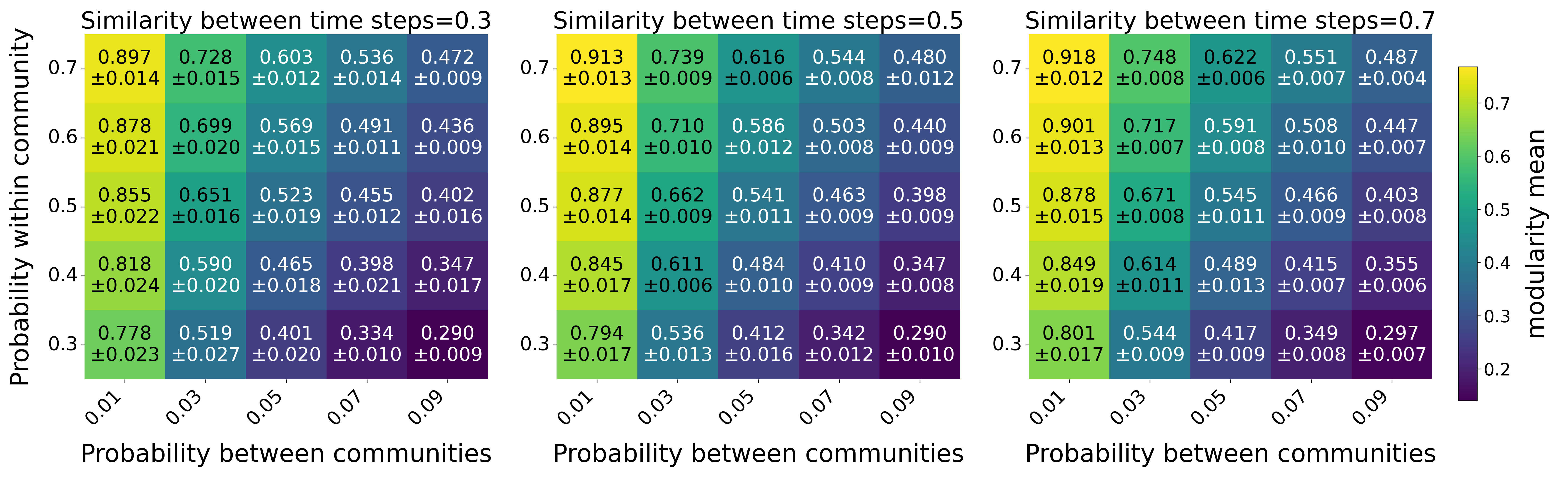}}
\subcaptionbox{Inter-layer persistence\label{fig:persistence}}{\includegraphics[width=0.5\linewidth]{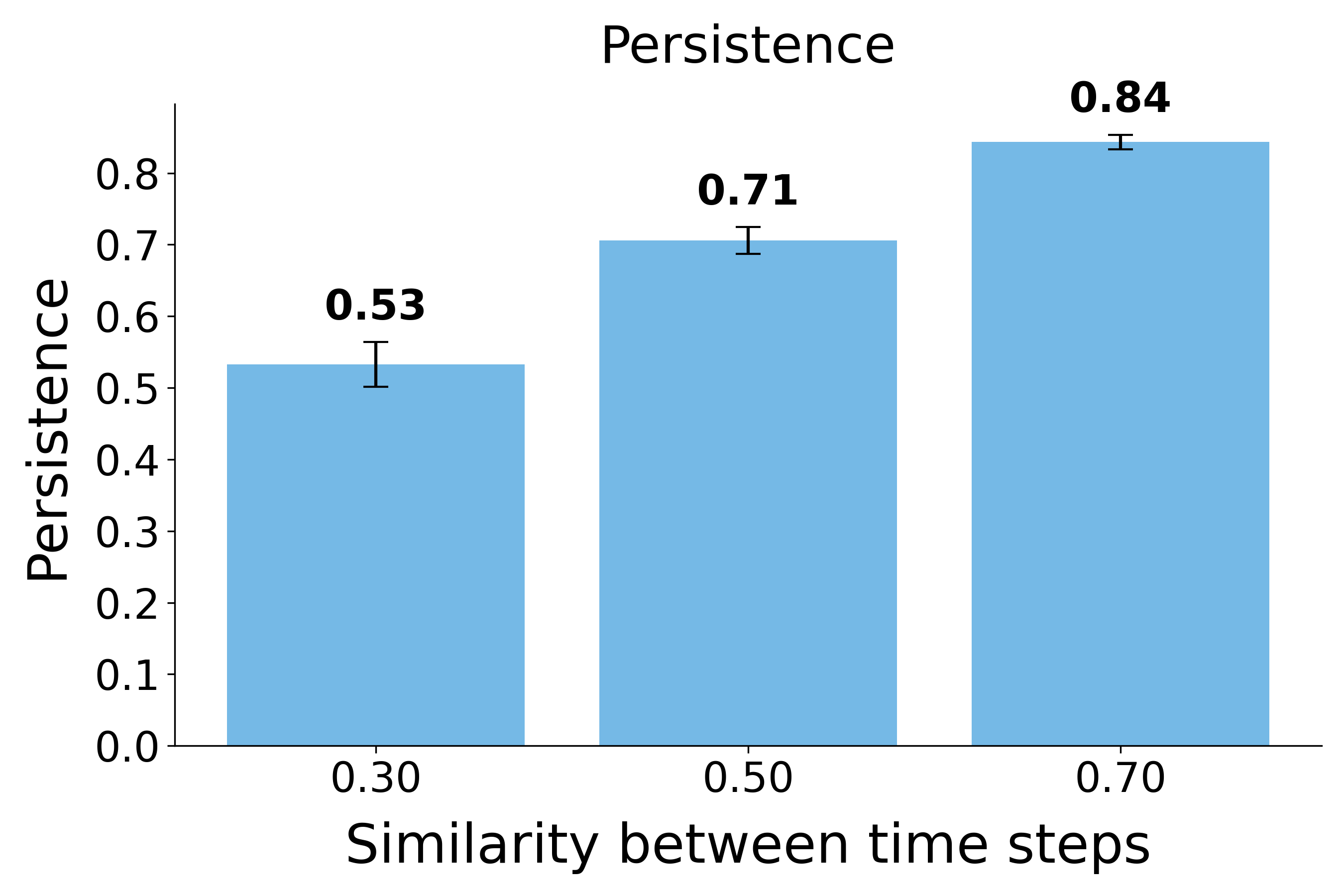}}
\caption{We consider multiple prescribed similarity levels between consecutive time steps, $s^{(t-1,t)}\in \{[0.3], [0.5], [0.7]\}$. For each time step, the probability of forming an edge between two nodes within the same community is $\pin^{(t)}=\{[0.3], [0.4], [0.5], [0.6], [0.7]\}$ and the probability of forming an edge between two nodes in different communities is \(\pout^{(t)} \in \{[0.01], [0.03], [0.05], [0.07], [0.09]\}\). To evaluate the quality of the synthetic community structure, we compute multilayer modularity, which captures both intra-snapshot structure and inter-snapshot consistency. (a) reports the modularity of individual snapshots, reflecting the strength of community structure at each time step; (b) reports the temporal persistence term, indicating how consistently nodes preserve their community labels across consecutive time steps.}
\label{fig:mm}
\end{figure}

We now demonstrate that the networks generated by the model exhibit community structures both in terms of time and topology. To quantify the quality of the community structure, we employ modularity, a commonly used measure of the strength of a network’s partition into communities. Because our model can generate dynamic networks both as a sequence of discrete snapshots and as a continuous link stream, we evaluate the quality of their community structure with two corresponding variants of modularity. The first variant is multilayer modularity~\cite{bazzi2016community,mucha2010community} treats each snapshot in time as a separate layer in a single, interconnected network. Within each layer, it measures the usual Newman–Girvan modularity by comparing the density of intra‐community edges to what would be expected at random. In addition, it introduces inter‐layer coupling terms that link each node to itself in adjacent time steps, enforcing temporal smoothness of community assignments. 

To adapt to the changing set of nodes, we introduce a node existence matrix $\alpha_i^t\in\{0,1\}$, to represent if node $i$ exists at time step $t$, where $\alpha_i^t=1$ represents node $i$ exists in time step $t$ and $\alpha_i^t=0$ means node $i$ does not exist in time step $t$. Then, the multilayer modularity when the node set changes is
\begin{equation}
\begin{aligned}
Q & =\frac{1}{|T|}\sum_{t=1}^{|T|} \sum_{i, j=1}^N \underbrace{\frac{1}{2\left|E_t\right|} \alpha_i^t \alpha_j^t M_{i j}^{(t)} \delta\left(c_i^t, c_j^t\right)}_{\text{Intra-layer modularity}} \\
& + \frac{1}{|T|-1}\sum_{t=1}^{|T|-1} \frac{1}{D_t}\sum_{i=1}^N \underbrace{\alpha_i^t \alpha_i^{t+1} \delta\left(c_i^t, c_i^{t+1}\right)}_{\text{Inter layer persistence}},
\end{aligned}
\label{eq:multilayer-mod}
\end{equation}
where $t$ denotes the index of the time step, and $D_t=\sum_{i=1}^N\alpha_i^{(t)}\alpha_i^{(t+1)}$. Furthermore, unlike the original form of multilayer modularity~\cite{mucha2010community}, we normalise intra-layer modularity and inter-layer persistence~\cite{bazzi2016community} to the range $[0, 1]$ respectively, thereby eliminating the bias of persistence caused by different $\pin^{(t)}$ and $\pout^{(t)}$. For each time step $t$, the modularity matrix is defined as $M^{(t)}=A^{(t)}-P^{(t)}$. $A^{(t)}$ is the adjacency matrix of the $t-$th time step, where $A_{ij}^{(t)}$ represents the number of edges connecting node $i$ and node $j$ at the $t$-th time step. $P_{ij}^{(t)}$ is the number of edges between the node $i$ and node $j$ in corresponding null model. The variable $c_i^{(t)}$ is the community label of node $i$ at time step $t$, and $\delta(\cdot,\cdot)$ is the Kronecker delta function. We use $N$ to represent the number of total nodes in the dynamic network, and there are $|T|$ layers and $|E_t|$ edges at time step $t$. In the sum of multilayer modularity, the first term accumulates contributions from all pairs of nodes $(i,j)$ that appear in the time step $t$, comparing the actual edge weight $A^{(t)}_{ij}$ with its expectation of the null model $P^{(t)}_{ij}$. The second term enforces temporal persistence by adding a contribution whenever node $i$ exists in both time steps $t$ and layer $t+1$ and retains the same community label across these adjacent layers. 

Figure~\ref{fig:mm} shows the multilayer modularity of the temporal network synthesised by the model under different combinations of parameters. We consider multiple prescribed similarity levels between consecutive time steps  $s^{(t-1,t)}=\{[0.3], [0.5], [0.7]\}$. For each time step, the probability of forming connections between nodes within the same community as $\pin^{(t)} = \{[0.3], [0.4], [0.5], [0.6], [0.7]\}$, and the probability of forming connections between nodes in different communities as $\pout^{(t)} = \{[0.01], [0.03], [0.05], [0.07], [0.09]\}$. The resulting temporal network is then segmented into multiple snapshots using a sliding window. Within each snapshot, the weight of an edge between two nodes is defined as the number of temporal interactions that occur between them during the window. To evaluate the impact of different parameter combinations on the quality of community structures of the synthesised temporal networks. Figure~\ref{fig:modularity} shows the average strength of the community structure in each time step, which corresponds to the first term of equation~\ref{eq:multilayer-mod}, namely, the modularity of the intra-layer. As the difference between $\pin^{(t)}$ and $\pout^{(t)}$ decreases, the number of edges within the community and the difference in the number of edges between communities decrease, thus making the structure of the community blurred. Figure~\ref{fig:persistence} shows the persistence of the community labels of the nodes over consecutive time steps, which corresponds to the second term of equation~\ref{eq:multilayer-mod}, namely inter-layer modularity. As the similarity of the community structure between adjacent time steps increases, the genetic algorithm has a higher probability of maintaining the community labels of the nodes, thereby increasing the persistence of the nodes. 
\subsection{Simulation}
\label{sec:simulation}
\begin{figure}
\centering
\includegraphics[width=\linewidth]{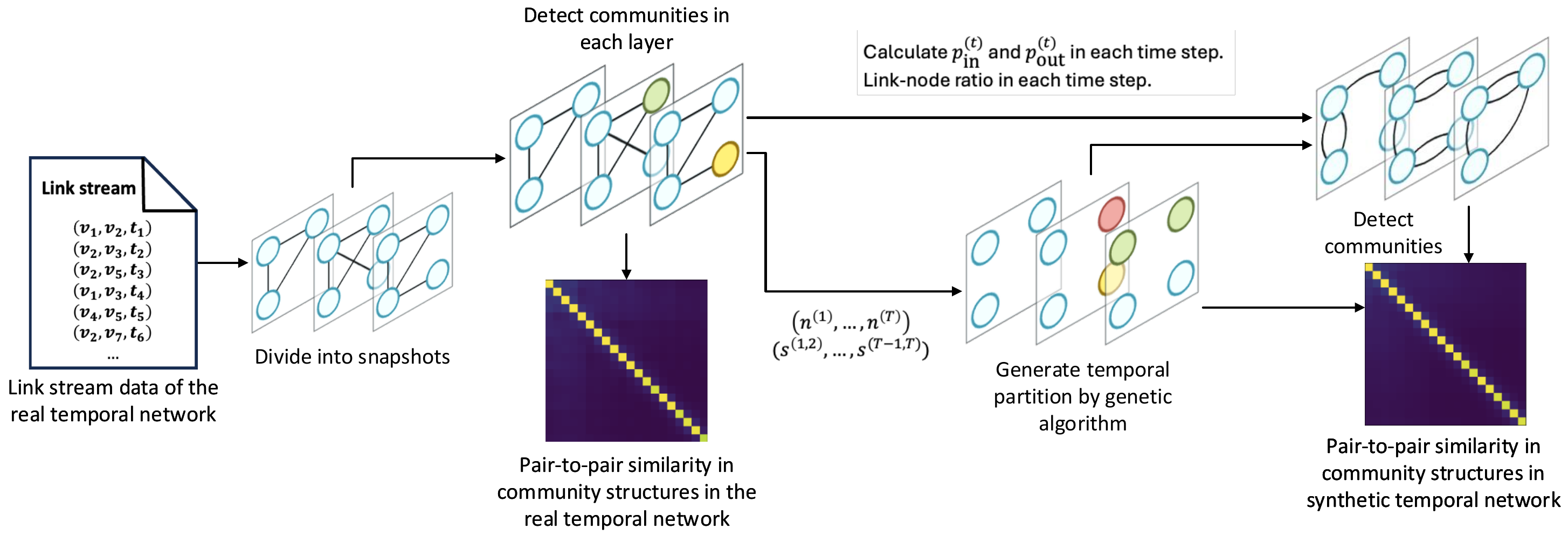}
\caption{This figure shows our workflow for synthesising a temporal network that simulates a real one. First, we apply a sliding-window approach and the Louvain algorithm to the real network to extract key parameters, including the number of nodes per snapshot and the step-to-step community structure similarity. These extracted parameters, along with connection probabilities ($\pin^{(t)}$, $\pout^{(t)}$) and the link-node ratio from the real data, are used to guide a genetic algorithm. This algorithm generates a new sequence of community partitions, and we then insert edges to create the synthetic network. Finally, we validate our model by applying the same community detection algorithm to the synthetic network and comparing its step-to-step community similarity with that of the original network.}
\label{fig:experiment-flow}
\end{figure}
To evaluate the capability of our dynamic network generation model to reproduce the structural features of real networks, we use the \texttt{email-Eu-core-temporal (Email), subReddit hyperlinks (SubReddit)~\cite{paranjape2017motifs}, high-energy physics theory citation network (HEP)~\cite{ryan2015the}}, and \texttt{non-fungible token (NFT) transaction~\cite{nadini2021mapping}} dataset. For the details of the datasets and preprocessing, refer to the appendix. The process of the experiment is demonstrated as Figure~\ref{fig:experiment-flow}. We first apply a sliding-window approach to divide the link stream data into multiple snapshots, where the weight of an edge within a snapshot corresponds to the number of interactions between a pair of nodes. Next, we employ the Louvain algorithm to detect the community structure at each snapshot and compute the step-to-step similarity of community structures in the empirical dynamic network. These parameters, the number of nodes in each snapshot and the similarity of community structures across consecutive snapshots are then provided as inputs to a genetic algorithm–guided community structure generator. Edges are subsequently inserted according to the the dynamic partitions generated by the genetic algorithm, and parameters  $p^{(t)}_{\mathrm{in}}$, $p^{(t)}_{\mathrm{out}}$, and the link–node ratio are extracted from the real data. Finally, we apply the same community detection algorithm to the generated network and compute the step-to-step similarity of its community structures for comparison.
\begin{figure}
    \centering
    \includegraphics[width=\linewidth]{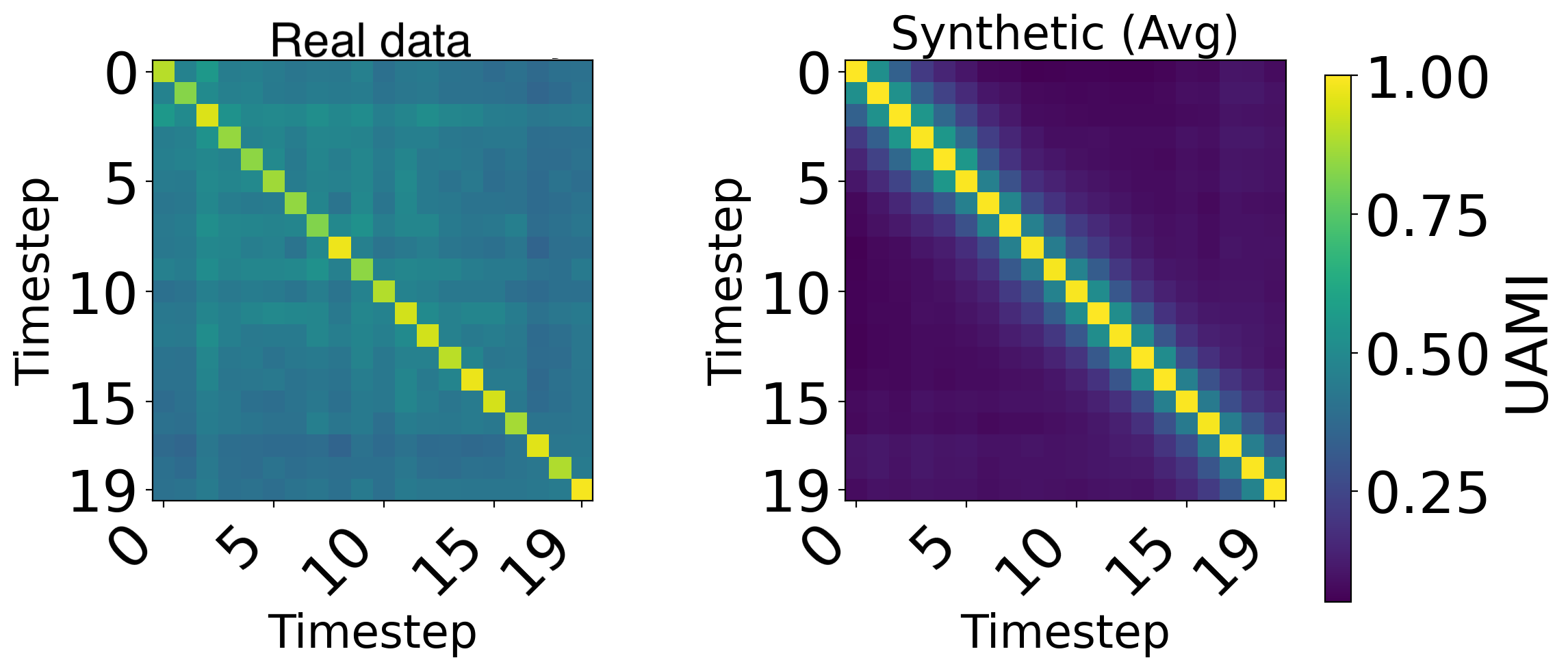}
    \caption{The figure demonstrates the similarity of community structures between pairs of time steps for both in a email communication temporal network and a set of synthetic networks.}
    \label{fig:email-simu}
\end{figure}

\begin{figure}
    \centering
    \includegraphics[width=\linewidth]{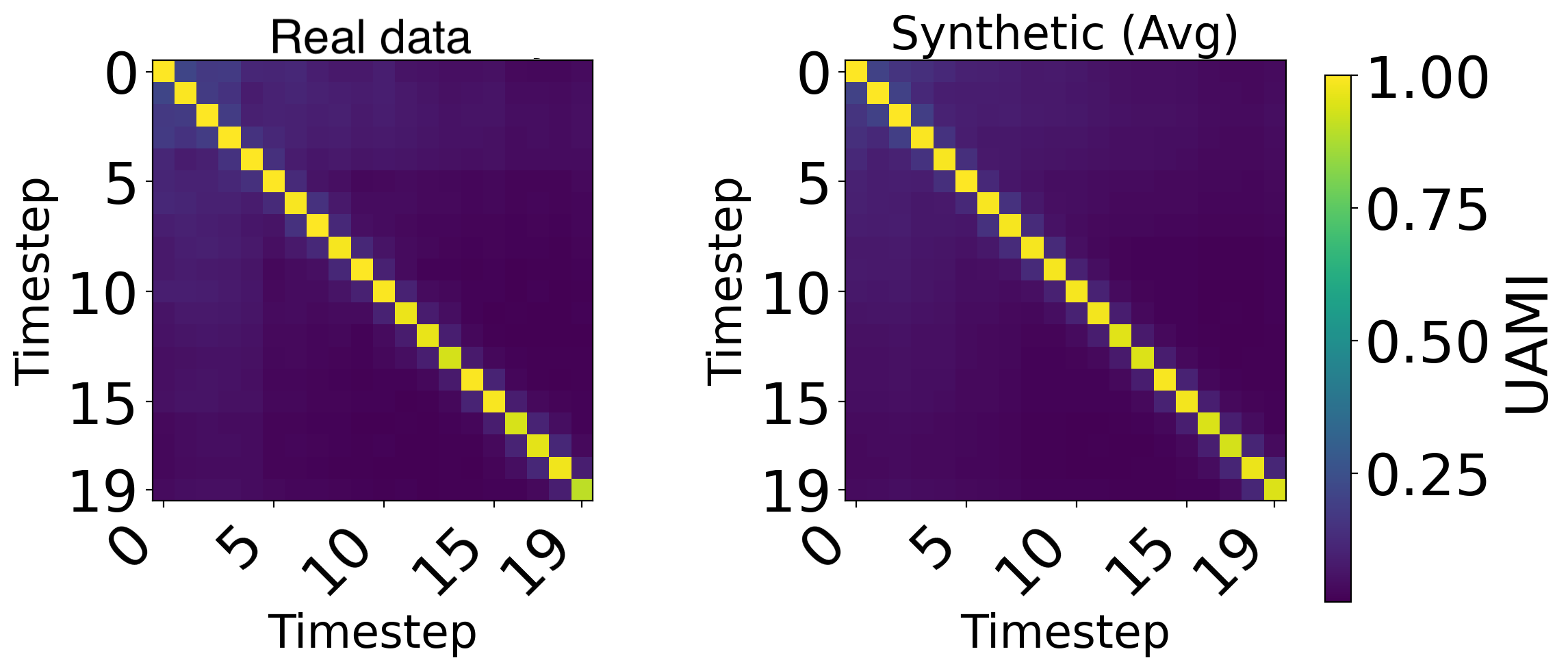}
    \caption{The figure demonstrates the similarity of community structures between pairs of time steps for both in a paper citation temporal network and a set of synthetic networks.}
    \label{fig:hep-simu}
\end{figure}

\begin{figure}
    \centering
    \includegraphics[width=\linewidth]{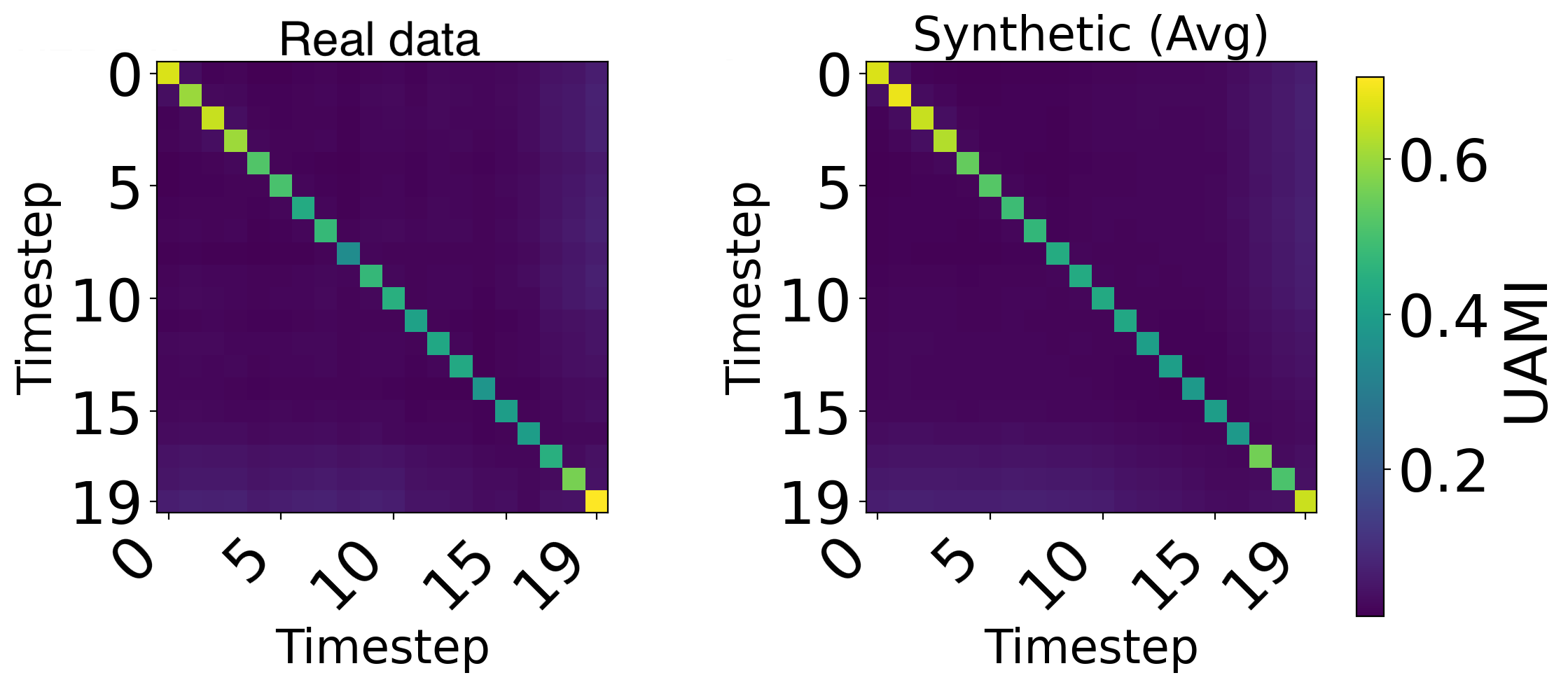}
    \caption{The figure demonstrates the similarity of community structures between pairs of time steps for both on a online social platform temporal network and a set of synthetic networks.}
    \label{fig:reddit-simu}
\end{figure}

\begin{figure}
    \centering
    \includegraphics[width=\linewidth]{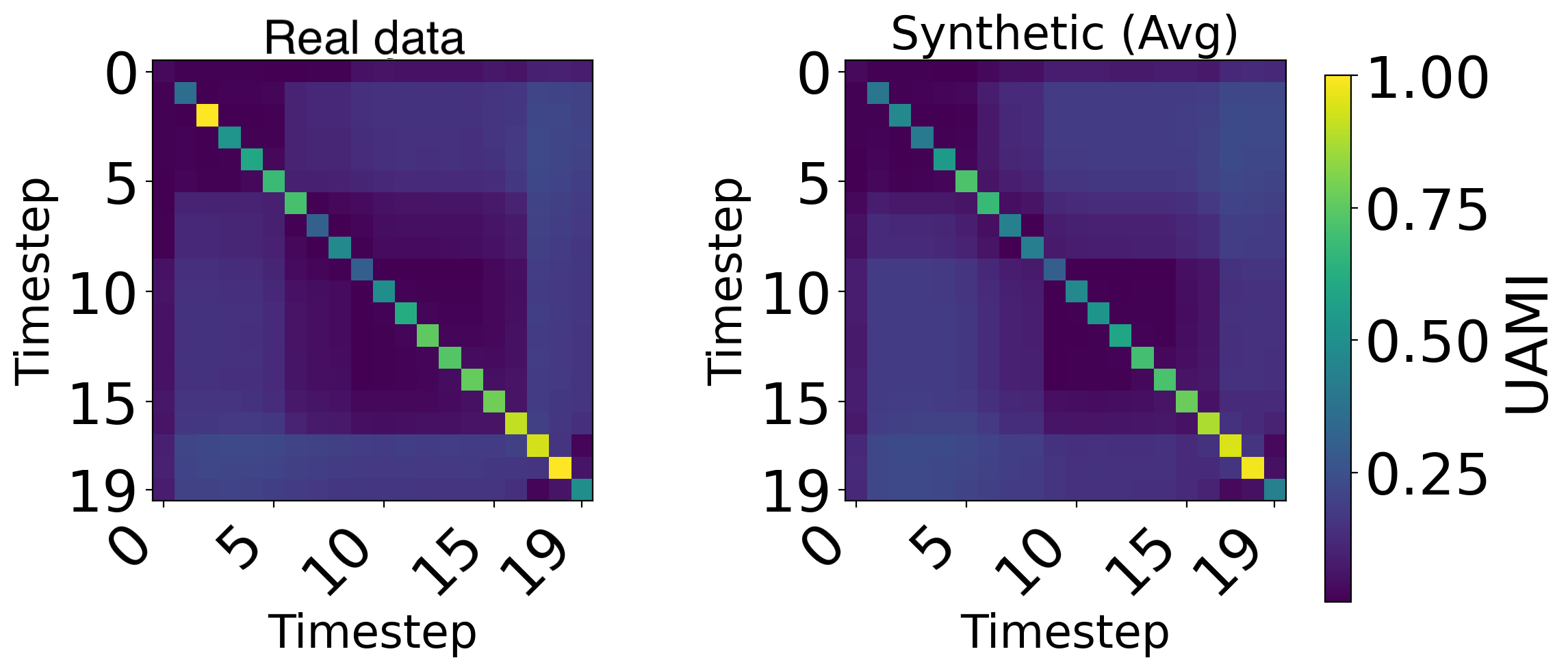}
    \caption{The figure demonstrates the similarity of community structures between pairs of time steps for both in a online transaction temporal network and a set of synthetic networks.}
    \label{fig:nft-simu}
\end{figure}

The core mechanism of our generative model is to optimise the similarity between the current and the next community structures to generate dynamic community partitions. Therefore, the dynamic communities generated by our model actually constitute a non-memory process. Therefore, we regarded this model as a null model to investigate the temporal dependence within real-world networks. Here, we test an assumption: the actual evolution of communities is more complex than a simple sequential process. Therefore, this empty model represents a baseline dynamic that lacks long-term structural memory. Any significant deviation of empirical data from this baseline will reveal the existence of higher-order temporal dynamics. 

From the overall experimental results, the model is able to effectively capture the community evolution characteristics of different types of dynamic networks. In the HEP and subreddit hyperlink data, the real network hardly has long-term memory, and the community structure is rapidly reorganised. The model accurately reproduces this characteristic. In the Email data, the real network shows the pattern of long-term stability of community structure and stable change rate, while the model successfully reproduces the short-range continuity. However, in terms of long-term similarity, it is slightly lacking. This is because of the model's non-memory feature. When synthesizing the network, the model only focuses on optimizing the similarity between adjacent time steps to reach the given target, but it is unable to simulate the long-term evolution characteristics of the community structure. In the NFT data, due to the frequent addition/removal of nodes from the network and the UAMI score increasing the long-term similarity through the virtual community mechanism when the node set varies greatly (refer to the appendix of the paper~\cite{zhong2024quantifying}), the real network shows the characteristic of short-term dissimilarity but relatively similar to the long-term community structure. The generative model still reproduces this phenomenon without explicitly modelling long-term dependencies. This indicates that the model is capable of not only modeling local dynamics well but also restoring the entry/exit of nodes from the network.
\subsection{Testbed for dynamic community detection}
\label{sec:test}

\begin{table}[t]
\centering
\caption{Summary of dynamic community detection algorithms}
\label{tab:algorithm_comparison}

\begin{tabular}{lcr}
\hline
\textbf{Algorithm} & \textbf{Strategy} & \textbf{Complexity} \\
\hline
GenLouvain~\cite{bazzi2016community}
    & Modularity maximisation
    & $O(nI^2L)$ \\

IterModMax~\cite{pamfil2019relating}
    & Modularity maximisation
    & $O(n^2L)$ \\

CLAN~\cite{dabideen2014clan}
    & Label propagation
    & $O(LInk)$ \\

NFC~\cite{aslak2018constrained}
    & Random walk
    & $O(m(n+m))$ \\
\hline
\end{tabular}

\vspace{0.5em}

\footnotesize
Comparison of algorithms with their strategies and complexities. Here, $n$ and $m$ denote the numbers of nodes and edges, respectively. $I$ denotes the number of iterations required to converge to a local optimum, and $L$ is the number of layers.

\end{table}

This generative model allows us to investigate how the performance of dynamic community algorithms is influenced by the rate at which nodes move in and out of the network, as well as the rate of change in the community structure. We modified DC-SBM~\cite{karrer2011stochastic} to generate edges with timestamps. Suppose that the community partition at time step $t$ is denoted as $\{C_1^{(t)},\dots,C_{k^{(t)}}^{(t)}\}$, and we estimate the minimum probability of connecting two nodes between two communities $\pout^{(t)}$ according to the method described in section~\ref{sec:insert-link}. We define a degree-correction sequence to tune the degree distribution over nodes $\Theta^{(t)}= \left(\theta^{(t)}_1, \dots, \theta^{(t)}_{n^{(t)}} \right)$, where $n^{(t)}$ is the number of nodes in the time step $t$. $\theta_v^{(t)}$ tunes the node degree of $v$ in time step $t$, satisfying the condition $\sum_{i=1}^{n^{(t)}} \theta^{(t)}_i=n^{(t)}$. A larger value of $\theta_{v}^{(t)}$ corresponds to a higher degree for node $v$ at that time step. To reflect the scale-free property observed in real-world networks, the sequence $\Theta^{(t)}$ is sampled from a power-law distribution in our experiments. 

Next, we introduce a parameter $\mu$ to control the degree of community separation, which is defined as $\mu=\frac{M^{(t)}_{\mathrm{inter}}}{M^{(t)}_{\mathrm{intra}}+M^{(t)}_{\mathrm{inter}}}$, where $M^{(t)}_{\mathrm{inter}}$ represents the number of edges connecting two nodes from different communities in time step $t$, and $M^{(t)}_{\mathrm{intra}}$ represents the number of edges connecting two nodes from the same community in time step $t$. A lower value of $\mu$ indicated stronger community separation, that is, fewer inter-community edges. From the definitions above, the number of potential inter-community edges can be given by $A_{\mathrm{inter}}^{(t)}=\sum_{1\leq r < s\leq k_t}\sum_{u\in C_r^{(t)},v\in C_s^{(t)}}\theta_u^{(t)}\theta_v^{(t)}$. Let $\pin^{(t)}$ be the probability of a connection between a pair of nodes within the same community at time step $t$, the number of potential intra-community edges is $A_{\mathrm{intra}}^{(t)}=\sum_{r}\sum_{u<v,\ u,v \in C_{r}^{(t)}}\theta_u^{(t)}\theta_v^{(t)}$. Thus, the total number of edges to be generated in the time step $t$ is $\pout^{(t)} A^{(t)}_{\mathrm{inter}}/\mu$, where the ratio of the number intra-community edges to the number of inter-community edges is given by $\mu/(1-\mu)$.
\begin{figure}
\centering
\subcaptionbox{The mAMI between dynamic community structure detected by NFC and the ground-truth community label.}{\includegraphics[width=\linewidth]{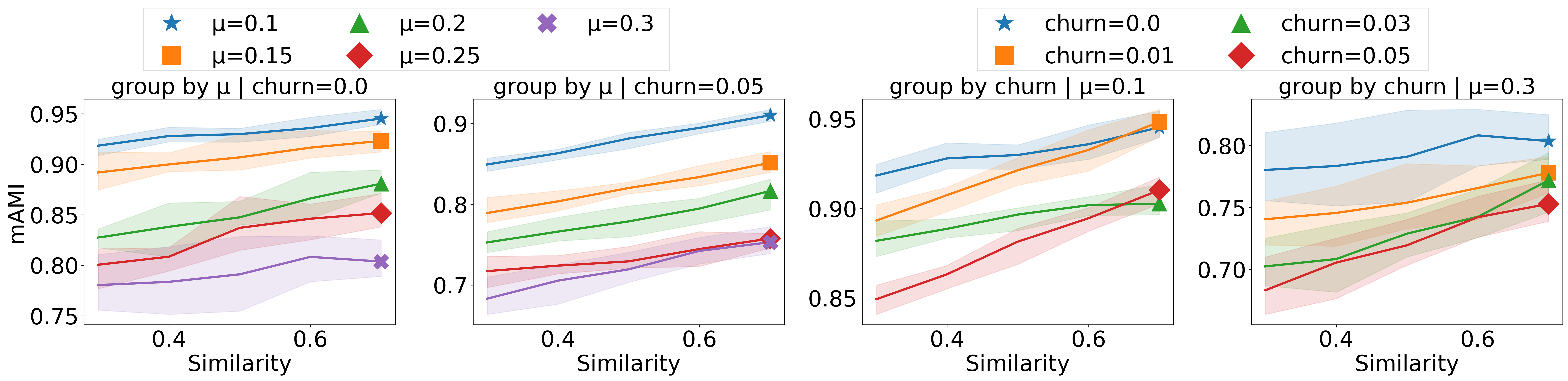}}
\subcaptionbox{
The mAMI between dynamic community structure detected by CLAN and the ground-truth community label.\label{fig:CLAN}
}{\includegraphics[width=\linewidth]{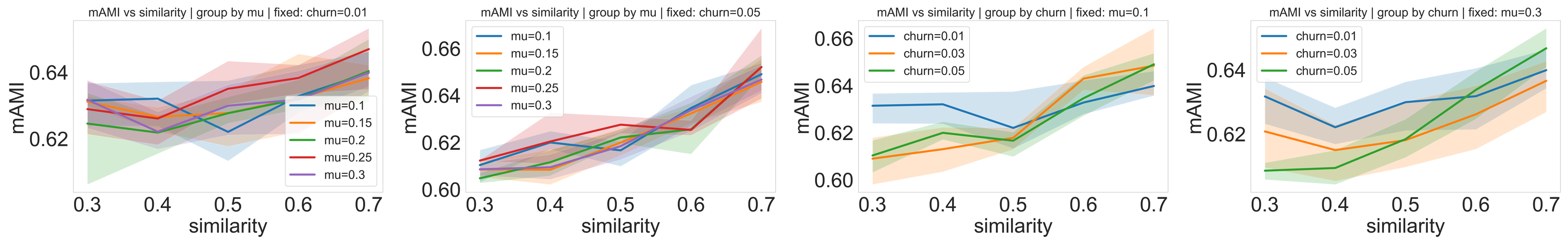}}
\subcaptionbox{
The mAMI between dynamic community structure detected by GenLouvain and the ground-truth community label.
}{\includegraphics[width=\linewidth]{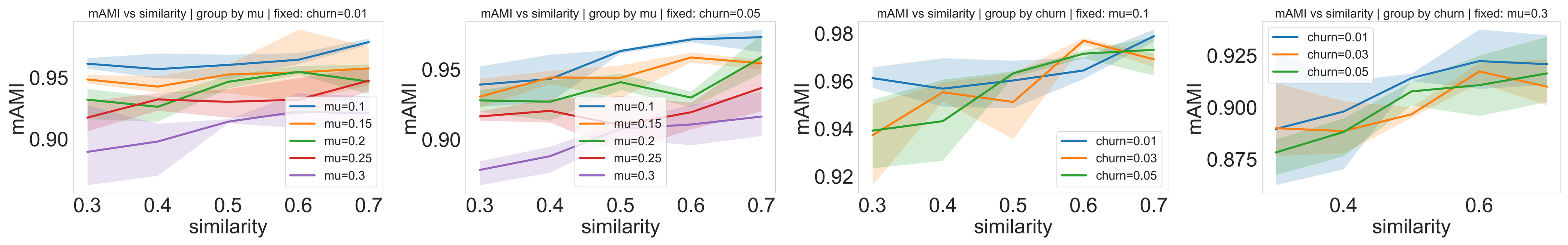}}
\caption{The figure illustrates the similarity between the detected community structures and the ground-truth labels for three different algorithms. The performance is evaluated under varying parameters: $\mu$ (which controls the degree of community separation); similarity (which controls the similarity of community structures between consecutive time steps); and churn (which dictates the rate at which nodes join and leave the network). The similarity is quantified by the mAMI score, where a value approaching one indicates a closer match between the detected and ground-truth community structures. The shaded area in the figure shows the range of the maximum and minimum values of mAMI in the community detection results of ten independently synthesised networks that have the same parameters set.}
\label{fig:test-algo}
\end{figure}

We synthesised dynamic networks with ten time steps, keeping a fixed size of 500 nodes at each step. To generate diverse network scenarios, we varied three key parameters: the similarity of the community structures between consecutive time steps (similarity), the degree of community separation ($\mu$), and the rate of nodes entering and leaving the network (churn). The performance of three algorithms: GenLouvain, NFC, and CLAN are evaluated on these synthesised networks. We use multiplex-adjusted mutual information (mAMI)~\cite{bazzi2020framework, aslak2018constrained} to measure the similarity between the true community labels and the community labels generated by the algorithm. It should be noted that when using the similarity measure based on mutual information (MI) to assess the similarity of community structures, what matters is the co-occurrence pattern of nodes within the same communities, rather than the numerical values of their community labels. For instance, one community assignment (1,1,2,2,2) and another (3,3,4,4,4) are equivalent in terms of MI, since the pairwise co-membership relationships among nodes are identical, yielding a mutual information value of one. Figure~\ref{fig:test-algo} reports the results of these experiments. 

The result shows, firstly, an increase in similarity significantly improves the mAMI score, indicating that when the community structure of adjacent time slices is more stable, the algorithm is more capable of leveraging historical information to maintain the consistency of community detection. Therefore, both NFC and GenLouvain can restore the true community very accurately when the similarity is high. Conversely, when the similarity is low, community evolution is intense, and the detection results are more likely to deviate from the true structure. Secondly, an increase in $\mu$ leads to a decrease in mAMI. This is because $\mu$ controls the degree of separation between communities. A larger $\mu$ means that the connections between communities are tighter and the boundaries are blurred more, making it difficult for the algorithm to distinguish communities, thus reducing the detection accuracy. Finally, there is a point that has been overlooked in previous studies: the increase in the rate of nodes entering and leaving the network also weakens the algorithm's performance. Because the frequent entry and exit of nodes disrupts the structural stability of the network and increases the uncertainty of detection; especially for NFC, this effect is more significant, indicating its insufficient robustness to dynamic node changes. In addition, CLAN shows consistently lower mAMI scores across almost all parameter settings (see Figure~\ref{fig:CLAN}). Its performance is generally worse than that of GenLouvain and NFC, suggesting that the label-propagation mechanism of CLAN is more sensitive to temporal perturbations and less capable of maintaining cross-time consistency in dynamic environments.
\subsection{Convergence tests}
\begin{figure}
\centering
\subcaptionbox{The initial partition with $n^{(1)}=400$ nodes is expanded to a target size of $\hat{n}^{(2)}=500$ nodes. This process is constrained to maintain a community structure similarity of 0.5. \label{fig:add-nodes}}{\includegraphics[width=\linewidth]{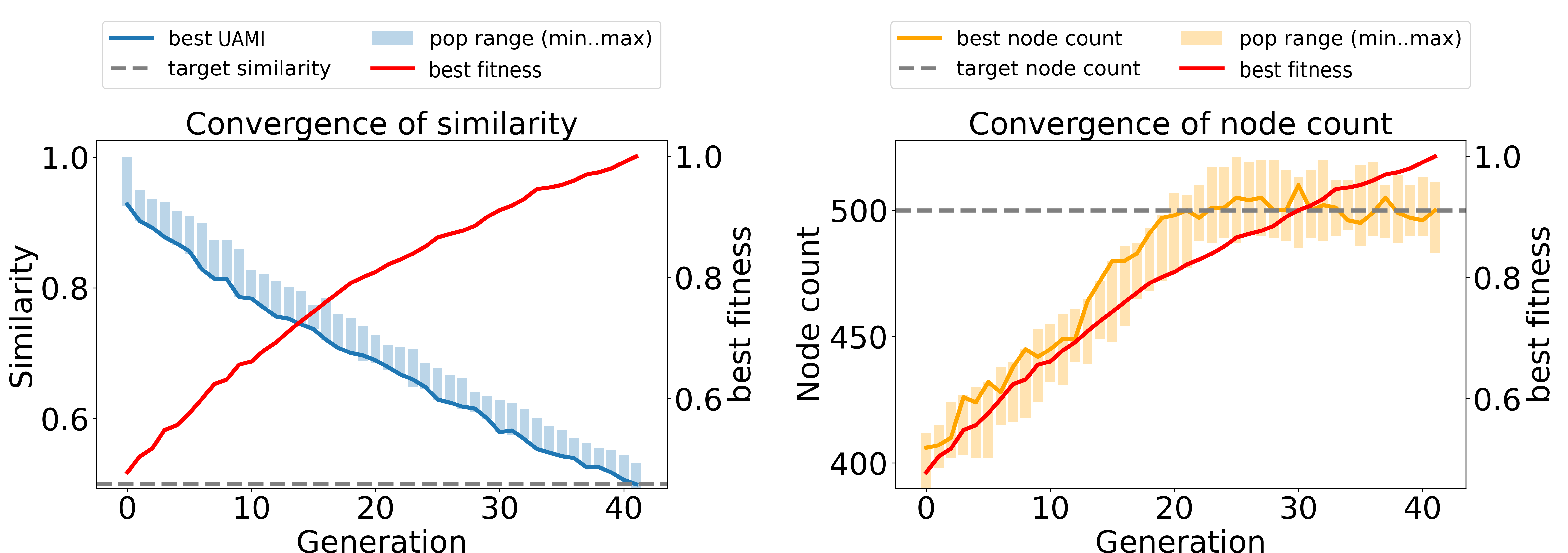}}
\subcaptionbox{The initial partition with $n^{(1)}=500$ nodes is shrank to a target size of $\hat{n}^{(2)}=400$ nodes. This process is constrained to maintain a community structure similarity of 0.5. \label{fig:remove-nodes}}{\includegraphics[width=\linewidth]{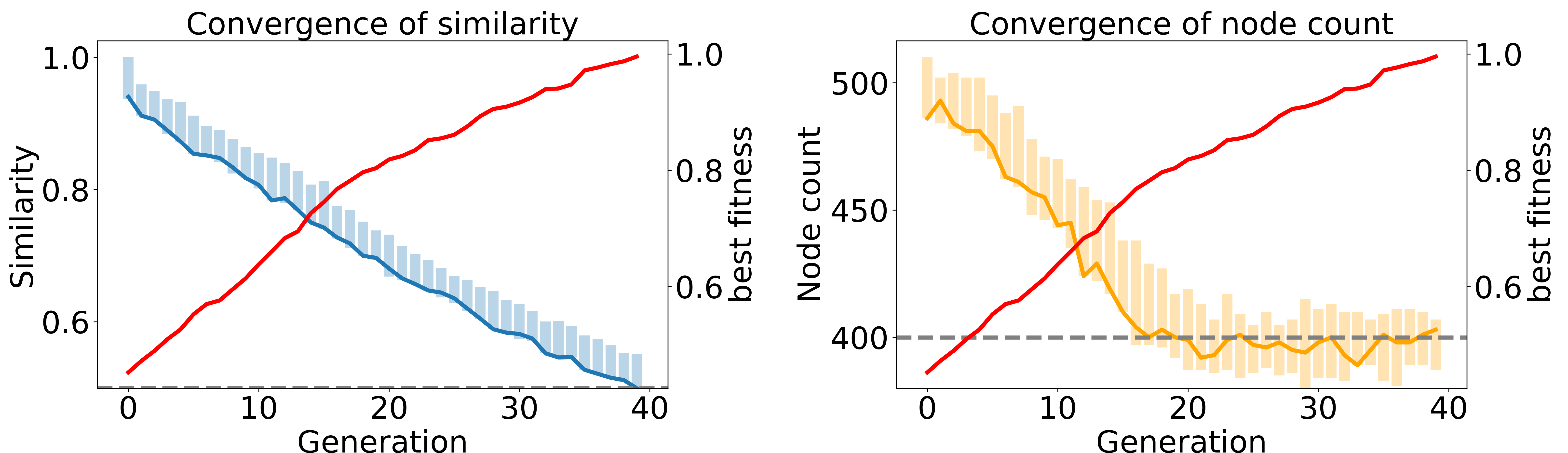}}
\caption{The model converges to the given community structure similarity and the number of nodes under different parameters. In the plot, best fitness refers to the individual with the highest fitness value in the current iteration. Best node count refers to the number of nodes of the best fit partition, and best UAMI refers to the similarity between the input partition and the best fit partition. }
\label{fig:convergence}
\end{figure}
To evaluate the convergence ability and search efficiency of the proposed genetic algorithm, we design a constrained optimisation experiment. The goal of this experiment is to verify whether the algorithm can discover a target partition in the solution space that satisfies a desired similarity to a input partition, and a desired number of nodes.

Specifically, the algorithm is initialised with a source partition $\mathcal{C}^{(1)}$, containing $n^{(1)}$ nodes. Its task is to produce a partition $\mathcal{C}^{(2)}$ with $\hat{n}^{(2)}$ nodes, while ensuring that the similarity between $\mathcal{C}^{(1)}$ and $\mathcal{C}^{(2)}$ converges to a prescribed value $s^{(1,2)}$. Figure~\ref{fig:convergence} reports the evolution of similarity and fitness across iterations under different mutation settings.
Figure~\ref{fig:add-nodes} illustrates the optimisation behaviour when nodes are added, while Figure~\ref{fig:remove-nodes} focuses on the case of node removal. In both settings, the fitness of the best individual increases steadily over time as the search progresses. Meanwhile, the similarity of the optimal solution gradually approaches the target value $s^{(1,2)}$ and eventually stabilises, indicating that the algorithm successfully enforces the structural constraint. As evolution proceeds, the population distribution shifts towards regions of higher fitness, demonstrating that the mutation and selection operators effectively guide the search in the solution space. Figure~\ref{fig:remove-nodes} further shows how the algorithm adapts to the constraint on the size of node count. As iterations increase, the number of nodes in candidate partitions becomes increasingly concentrated around the desired value $\hat{n}^{(2)}$. Most high-fitness individuals either exactly satisfy or closely approximate $\hat{n}^{(2)}$, confirming that the algorithm can simultaneously maintain structural similarity and achieve the correct scale.
\section{Discussion}
We have presented a comprehensive two-stage generative model for temporal networks that simultaneously controls community evolution and dynamic node sets. The key innovation lies in the utilisation of a genetic algorithm guided by our measure, union-adjusted mutual information, to generate sequences of communities. The model enables explicit control over inter-snapshot similarity and complex events like node churn, while ensuring network connectivity through control over link density between communities. Experimental evaluations demonstrate that the model replicates short-term evolutionary patterns in real data and also provides a benchmark that allows researchers to assess the robustness of community detection algorithms where there is node churn and where temporal communities may evolve at different rates. 

Despite its flexibility, the current model relies on a first-order Markovian assumption: that the communities in time slice $t$ rely only on those in time slice $t-1$. While this memoryless property serves as an excellent null model to detect higher-order dependencies in real data, it may not fully capture systems with long-term memory or bursty dynamics, where past events significantly influence future structural changes. An example with long-term memory can be seen in \texttt{email-EU-core} network where the network has considerable memory as it contains long-term collaborators working in the same organisation. Future iterations of this framework could incorporate a memory module into the genetic algorithm, where the fitness function considers a window of past partitions rather than just the immediate predecessor. Additionally, while the current edge generation uses a Poisson process for timestamps, integrating self-exciting point processes, such as Hawkes processes, could better model the temporal clustering of interactions often observed in human communication networks. Finally, exploring parallelisation strategies for the genetic algorithm could further enhance scalability, enabling the synthesis of large-scale temporal networks with millions of nodes.

In summary, this work bridges the gap between static graph community generators and the complex reality of evolving systems. By providing a controllable, interpretable, and theoretically grounded tool for generating temporal networks with dynamic communities and node churn, we offer the complex networks researchers a powerful instrument to advance the understanding of temporal network dynamics and the development of more robust analysis algorithms.

\bibliographystyle{plain}
\bibliography{ref}
\end{document}